# Optimized Resource Allocation in Multi-user WDM VLC Systems

**Sarah O. M. Saeed[1], Sanaa Hamid Mohamed[1], Osama Zwaid Alsulami[1], Mohammed T. Alresheedi[2], and Jaafar M. H. Elmirghani[1]**

[1]*School of Electronic and Electrical Engineering, University of Leeds, LS2 9JT, U.K.*
[2]*Department of Electrical Engineering, King Saud University, Riyadh, Kingdom of Saudi Arabia*
elsoms@leeds.ac.uk, elshm@leeds.ac.uk, ml15ozma@leeds.ac.uk, malresheedi@ksu.edu.sa, j.m.h.elmirghani@leeds.ac.uk

**ABSTRACT**
In this paper, we address the optimization of wavelength resource allocation in multi-user WDM Visible Light Communication (VLC) systems. A Mixed Integer Linear Programming (MILP) model that maximizes the sum of Signal-to-Interference-plus-Noise-Ratio (SINR) for all users is utilized. The results show that optimizing the wavelength allocation in multi-user WDM VLC systems can reduce the impact of the interference and improve the system throughput in terms of the sum of data rates for up to 7 users.

## 1. INTRODUCTION

The demands for high data rates are expected to dramatically increase as Internet-connected devices are expected to be more than half of the global population by 2022 generating an increased global IP traffic by three folds [1]. Hence, enhanced and future-proof technologies are needed to continue fulfilling these future demands. The radio spectrum has already become congested and to cope with the increasing bandwidth requirements, it needs to be supplemented by frequencies that can support high data rates such as infrared, visible, and ultraviolet spectrum technologies which are used in optical wireless communication (OWC) systems. While OWC has attracted the attention of researchers for more than three decades [2], Visible Light Communication (VLC) has gained more interest as the communication can be a by-product of existing illumination infrastructure. The use of Solid-State Lighting (SSL) for indoor illumination has increased over the past decade and is expected to provide 75% of the aggregate global lighting by 2030 [3]. SSL devices are characterised by longer lifespan, and energy efficiency in addition to their ability to switch between different light intensity levels at a very fast rate [4]. The last property is a unique one compared to conventional lighting technologies, which can be exploited in communication to encode data.

While VLC systems have the potential to provide high data rates, indoor VLC channels suffer from multipath propagation which causes Inter-Symbol Interference (ISI) and Co-Channel Interference (CCI) in systems with multiple users. Both, ISI and CCI result in decreasing the data rate of the VLC system. Changing the number of transmitters and receivers, in addition to controlling their directionality creates several favourable configurations for VLC systems that can minimize the dispersion resulting from multipath propagation and increase the SNR at the receiver [5]–[9].

Many studies aimed at increasing the achievable data rate and previous work showed that multi-gigabit per seconds could be obtained [10]–[16] . In [10]-[12] multi-beam power adaptation with imaging receivers was used, while in [13], the beam angle and beam power were adapted with reception using an Angle Diversity Receiver (ADR). Delay adaptation with imaging diversity receiver was used in [14], and also in [15], [16] with the addition of relay nodes. Holograms to steer the whole beam power to the user's location are used in [17]–[20]. The adaptation techniques used in [10]–[16] enabled the achievement of up to 10 Gbps data rate at an increase in the computational cost. To avoid this cost, a Fast Computer-Generated Hologram (FCGH) algorithm was used in [17] and [21] and reported in [17] to achieve up to 25 Gb/s. Line strip multi-beam spot diffusing transmitters were studied in [22], [23] as a means to achieve high data rates in the absence of line of sight components. The problem of glare due to VLC uplink was tackled in [24] introducing an infrared uplink design in VLC.

The previous studies outlined above considered a single-user scenarios. Optimizing the allocation of a transmitter to a user in a single user scenario was done in [25] using Sub-Carrier Multiplexing (SCM) while considering the mobility of the single user. The use of SCM in [25] was extended to a multi-user scenario in [26] using multi-branch transmitters and also in [27] with Wavelength Division Multiplexing (WDM). In multi-user scenarios, the interference between users leads to degradation in the signal quality, hence, multiple access optimization techniques that allow efficient use of different resources such as wavelength, space, time, and power, have recently received increased attention [28].

In this paper, we optimize the allocation of wavelengths and multiple access points to users in WDM VLC systems by using a Mixed Integer Linear Programming (MILP) model that maximizes the sum of Signal to Interference-plus-Noise-Ratio (SINR) for all users. The rest of this paper is organized as follows: The system model and the parameters used are introduced in Section 2. The results are provided in Section 3 while the conclusions and future work are provided in Section 4.

## 2. SYSTEM MODEL

In this work, we consider an unfurnished room with no doors or windows and a (length × width × height) dimensions of (8 m × 4 m × 3 m) as shown in Figure 1. Eight lighting units (also access points) are used to provide adequate illumination in the room according to the EU and ISO standards [14], [15]. To achieve high modulation bandwidth, Laser Diodes (LDs) are used and the white illumination is obtained using four-colours: Red, Yellow, Green, and Blue (RYGB) LDs as in [14], [15], [17], [18], [29]. Lambertian radiation with 70° half power semi-angle is assumed to ensure the room is well lit. Users are randomly distributed over the communication floor which is a plane 1 m above the ground (i.e. the height of a typical office desk). We considered a wide Field Of View (wFOV) receiver that uses On-Off Keying (OOK) modulation; and for simplicity, we only considered the Line-Of-Sight (LOS) component. We assumed receivers with equal wavelength responsivities and with tuneable filters. Therefore, when a user is assigned to a wavelength, all other wavelengths can be ignored. The key parameters considered are summarized in Table 1.

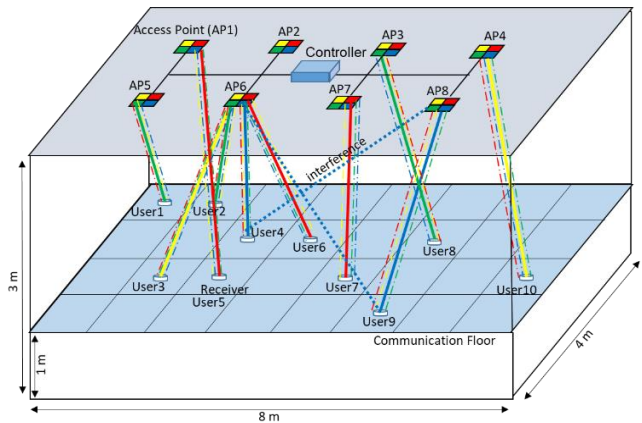

Figure 1. *Room configuration with 8 access points and 10 users distributed over the communication floor with random allocation of wavelengths. A solid line indicates that an allocated wavelength is used for communication. A dot-dashed line indicates that a wavelength is used for illumination only. A dotted line indicates interference between communication channels which is illustrated for the blue wavelength only.*

For such a system, the SINR for the $i^{th}$ user can be given by [21]:

$$SINR_i = \frac{(R.P_t.h_i)^2}{\overline{\sigma_t^2} + \sum_i Interference_i} \quad (1)$$

where $R$ is the responsivity of the photodetector in (A/W), $P_t$ is the transmitted optical power, $h_i$ is the optical channel gain for the $i^{th}$ user, between the user and the serving access point, $Interference_i$ is the interference received by the $i^{th}$ user from other access points with the same wavelength, and $\sigma_t$ is the receiver's total noise standard deviation given by:

$$\sigma_t = \sqrt{\overline{\sigma_{bn}^2} + \overline{\sigma_s^2} + \overline{\sigma_{pr}^2}} \quad (2)$$

where $\sigma_{bn}, \sigma_s,$ and $\sigma_{pr}$ represent the background light shot noise, shot noise associated with the signal, and the preamplifier noise, respectively calculated as in [30]. As stated in the previous section, a receiver can tune to a specific wavelength, hence, the background light shot noise is considered only for that wavelength. We considered the PIN-FET receiver proposed by Kimber et. el. [31] with noise spectral density of 10pA/√Hz and a bandwidth of 7 GHz.

*Table 1. Key parameters for the system model.*

| Parameter | Value |
|---|---|
| Room dimensions: width×length×height | 4m×8m×3m |
| Communication floor height | 1m |
| Number RYGB LDs lighting units | 8 |
| Location of lighting units | (1 m, 1 m, 3 m), (1 m, 3 m, 3 m), (1 m, 5 m, 3 m), (1 m, 7 m, 3 m), (3 m, 1 m, 3 m), (3 m, 3 m, 3 m), (3 m, 5 m, 3 m), (3 m, 7 m, 3 m) |
| Transmitted powers for RYGB colour | 800, 500, 300, 300 mW |
| Half-power semi-angle | 70° |
| Photodetector area | 1 cm$^2$ |
| Photodetector Responsivity | 0.4 A/W |
| Receiver noise current density | 10 pA/√Hz |
| Receiver electrical bandwidth | 7 GHz |

We utilized a MILP model to optimize the allocation of the wavelengths (R, Y, G, and B) and access points (1 to 8) to users randomly distributed over the communication floor in a Poison Point Process (PPP). The model maximizes the total SINR by maximizing the sum of the received optical signal power for all users and minimizing the sum of the receiver's noise for all users and the sum of the interference for all users. The model ensures that all users are served and that each user is allocated to a unique combination of wavelength and access point. It is assumed that the allocation is done by a centralized controller (see Figure 1) that has prior knowledge of the users' locations and their received power from all access points.

## 3. RESULTS AND DISCUSSION

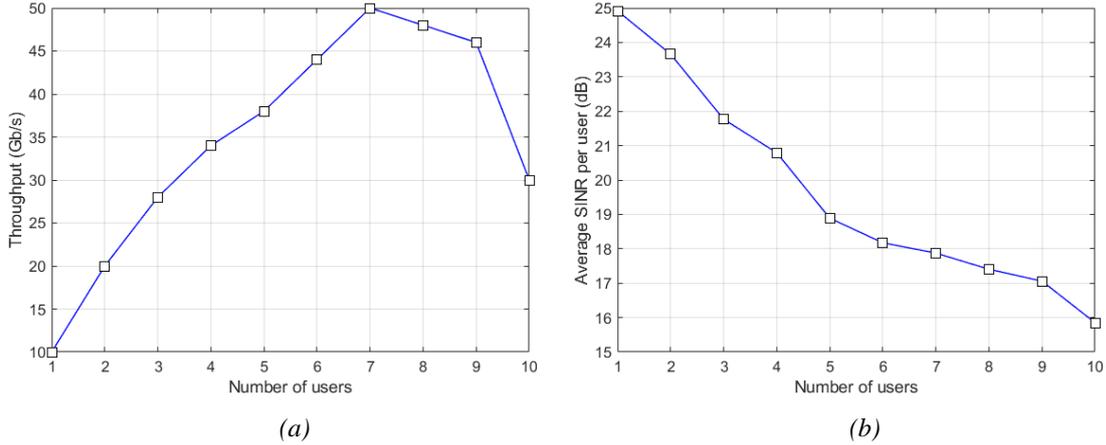

*Figure 2. (a)Throughput (sum of rates) in Gb/s vs number of users. (b) Average SINR per user in dB vs number of users.*

Figure 2 shows the results for the overall system throughput and average SINR per user. Each point in the curves is obtained by averaging the results for five random locations of users over the communication floor generated using PPP. Figure. 2 (a) shows the sum of the achievable data rates at a BER of 10$^{-9}$, where users that could not achieve this BER are not considered. It was noticed that for up to two users, each user can achieve 10 Gb/s. As the number of users increases beyond two users, the throughput increases at a lower rate as a result of the increase in the interference. The maximum throughput is achieved when the number of users in the room is 7 for the set of parameters considered, and depending on the random locations of users, mostly, three of the wavelengths are used twice and one wavelength is used once (i.e., this wavelength has zero interference). As the number of users increases and the distance between the users decreases, the impact of the interference becomes more significant resulting in decrease in the achieved system throughput, where the throughput decreases at a faster rate. This can be attributed to the fact that serving all users results in increase in the use of the same wavelengths. Figure 2 (b) shows the decrease in the average SINR per user as the number of users increases, which is also attributed to the increase of the interference.

## 4. CONCLUSIONS AND FUTURE WORK

In this paper, we optimized the wavelength allocation in a multi-user WDM VLC system with wFOV receivers using a MILP model that maximizes the sum of SINR for all users. The results show an overall increase in the system throughput for up to seven users. However, increasing the number of users beyond that results in a reduction in the achievable throughput. Future work includes the consideration of the first and the second order reflections of the transmitted optical power. This will account for multipath dispersion and will also account for situations where no LOS power component is available. Furthermore, to reduce the impact of the interference in WDM VLC systems, cell size optimization, in addition to Time Division Multiplexing (TDM) and different transmitter and receiver architectures can be considered.

**ACKNOWLEDGEMENTS**

The authors would like to acknowledge funding from the Engineering and Physical Sciences Research Council (EPSRC) for the TOWS project (EP/S016570/1). The first author would like to acknowledge her PhD funding from the School of Electronic and Electrical Engineering, University of Leeds, UK and from the Ministry of Higher Education and Scientific Research, Sudan. All data are provided in full in the results section of this paper.

**REFERENCES**

1. Cisco, "Cisco Visual Networking Index: Forecast and Trends, 2017–2022," 2019.
2. F. R. Gfeller and U. Bapst, "Wireless In-House Data Communication via Diffuse Infrared Radiation," *Proc. IEEE*, vol. 67, no. 11, pp. 1474–1486, 1979.
3. G. Zissis, *2014 Update on the Status of LED market*. 2014.
4. P. H. Pathak, X. Feng, P. Hu, and P. Mohapatra, "Visible Light Communication, Networking, and Sensing: A Survey, Potential and Challenges," *IEEE Commun. Surv. Tutorials*, vol. 17, no. 4, pp. 2047–2077, 2015.
5. K. L. Sterckx, J. M. H. Elmirghani, and R. A. Cryan, "Sensitivity assessment of a three-segment pyramidal fly-eye detector in a semidisperse optical wireless communication link," *IEE Proc. Optoelectron.*, vol. 147, no. 4, pp. 286–294, 2000.
6. K. L. Sterckx, "Pyramidal fly-eye detection antenna for optical wireless systems," *Digest IEE Colloq. on Optical Wireless Communications, Digest No. 1999*, p. 5/1-5/6.
7. A. G. Al-Ghamdi and M. H. Elmirghani, "Optimization of a triangular PFDR antenna in a fully diffuse OW system influenced by background noise and multipath propagation," *IEEE Trans. Commun.*, vol. 51, no. 12, pp. 2103–2114, 2003.
8. A. G. Al-Ghamdi and J. M. H. Elmirghani, "Characterization of mobile spot diffusing optical wireless systems with diversity receiver," vol. 00, no. c, p. 133–138 Vol.1, 2004.
9. A. G. Al-Ghamdi and J. M. H. Elmirghani, "Performance evaluation of a triangular pyramidal fly-eye diversity detector for optical wireless communications," *IEEE Commun. Mag.*, vol. 41, no. 3, pp. 80–87, 2003.
10. F. E. Alsaadi and J. M. H. Elmirghani, "Performance evaluation of 2.5 Gbit/s and 5 Gbit/s optical wireless systems employing a two dimensional adaptive beam clustering method and imaging diversity detection," *IEEE J. Sel. Areas Commun.*, vol. 27, no. 8, pp. 1507–1519, 2009.
11. F. E. Alsaadi and J. M. H. Elmirghani, "Mobile Multigigabit Indoor Optical Wireless Systems Employing Multibeam Power Adaptation and Imaging Diversity Receivers," *J. Opt. Commun. Netw.*, vol. 3, no. 1, pp. 27-39, 2010.
12. M. T. Alresheedi and J. M. H. Elmirghani, "10 Gb/s indoor optical wireless systems employing beam delay, power, and angle adaptation methods with imaging detection," *IEEE/OSA J. Light. Technol.*, vol. 30, no. 12, pp. 1843–1856, 2012.
13. M. T. Alresheedi and J. M. H. Elmirghani, "Performance evaluation of 5 Gbit/s and 10 Gbit/s mobile optical wireless systems employing beam angle and power adaptation with diversity receivers," *IEEE J. Sel. Areas Commun.*, vol. 29, no. 6, pp. 1328–1340, 2011.
14. A. T. Hussein and J. M. H. Elmirghani, "Mobile Multi-Gigabit Visible Light Communication System in Realistic Indoor Environment," *J. Light. Technol.*, vol. 33, no. 15, pp. 3293–3307, 2015.
15. A. T. Hussein and J. M. H. Elmirghani, "10 Gbps Mobile Visible Light Communication System Employing Angle Diversity, Imaging Receivers, and Relay Nodes," *J. Opt. Commun. Netw.*, vol. 7, no. 8, pp. 718-735, 2015.
16. F. E. Alsaadi, M. Nikkar, and J. M. H. Elmirghani, "Adaptive mobile optical wireless systems employing a beam clustering method, diversity detection, and relay nodes," *IEEE Trans. Commun.*, vol. 58, no. 3, pp. 869–879, 2010.
17. A. T. Hussein, M. T. Alresheedi, and J. M. H. Elmirghani, "Fast and Efficient Adaptation Techniques for Visible Light Communication Systems," *J. Opt. Commun. Netw.*, vol. 8, no. 6, p. 382, 2016.
18. A. T. Hussein, M. T. Alresheedi, and J. M. H. Elmirghani, "20 Gb/s Mobile Indoor Visible Light Communication System Employing Beam Steering and Computer Generated Holograms," *J. Light. Technol.*, vol. 33, no. 24, pp. 5242–5260, 2015.
19. M. T. Alresheedi and J. M. H. Elmirghani, "Hologram Selection in Realistic Indoor Optical Wireless

Systems With Angle Diversity Receivers," *J. Opt. Commun. Netw.*, vol. 7, no. 8, pp. 797-813, 2015.
20. M. T. Alresheedi, A. T. Hussein, and J. M. H. Elmirghani,, "Holograms in Optical Wireless Communications," *Opt. Fiber Wirel. Commun, R. Róka, Ed. InTech, 2017, pp. 125–141.*, vol. i, p. 13, 2018.
21. F. E. Alsaadi, M. A. Alhartomi, and J. M. H. Elmirghani, "Fast and efficient adaptation algorithms for multi-gigabit wireless infrared systems," *J. Light. Technol.*, vol. 31, no. 23, pp. 3735–3751, 2013.
22. A. G. Al-Ghamdi and J. M. H. Elmirghani, "Line Strip Spot-Diffusing Transmitter Configuration for Optical Wireless Systems Influenced by Background Noise and Multipath Dispersion," *IEEE Trans. Commun.*, vol. 52, no. 1, pp. 37–45, 2004.
23. F. E. Alsaadi and J. M. H. Elmirghani, "Adaptive mobile line strip multibeam MC-CDMA optical wireless system employing imaging detection in a real indoor environment," *IEEE J. Sel. Areas Commun.*, vol. 27, no. 9, pp. 1663–1675, 2009.
24. M. T. Alresheedi, A. T. Hussein, and J. M. H. Elmirghani, "Uplink design in VLC systems with IR sources and beam steering," *IET Commun.*, vol. 11, no. 3, pp. 311–317, 2017.
25. S. H. Younus, A. A. Al-Hameed, A. T. Hussein, M. T. Alresheedi, and J. M. H. Elmirghani, "Parallel Data Transmission in Indoor Visible Light Communication Systems," *IEEE Access*, vol. 7, pp. 1126–1138, 2019.
26. S. H. Younus, A. A. Al-hameed, A. T. Hussein, M. T. Alresheedi, and J. M. H. Elmirghani, "Multi-branch Transmitter for Indoor Visible Light Communication Systems," pp. 1–12, published in ResearchGate, November 2018, DOI: 10.13140/RG.2.2.29677.64483, arXiv:1812.06928, 12 Nov 2018.
27. S. H. Younus, A. A. Al-Hameed, A. T. Hussein, M. T. Alresheedi, and J. M. H. Elmirghani, "WDM for Multi-user Indoor VLC Systems with SCM," pp. 1–11, arXiv:1811.01341, and Researchgate, Nov. 2018.
28. M. Obeed, A. M. Salhab, M.-S. Alouini, and S. A. Zummo, "On Optimizing VLC Networks for Downlink Multi-User Transmission: A Survey," *IEEE Commun. Surv. Tutorials*, vol. PP, no. c, p. 1, 2018.
29. W. Davis, J. Y. Tsao, Y. Ohno, S. R. J. Brueck, J. J. Wierer, and A. Neumann, "Four-color laser white illuminant demonstrating high color-rendering quality," *Opt. Express*, vol. 19, no. S4, p. A982, 2011.
30. J. M. Kahn and J. R. Barry, "Wireless infrared communications," *Proc. IEEE*, vol. 85, no. 2, pp. 265–298, 2002.
31. E. M. Kimber, "High perfromance 10 Gbit/s pin-FET optical receiver," vol. 28, no. 2, pp. 16–18, 1992.